\documentstyle[11pt]{article}
\def \d {{\rm d}}

\newcommand{\A}{{\cal A}}
\newcommand{\M}{{\cal M}}

\begin{document}

\title{Exact non-singular  waves in the anti--de Sitter universe}

\author{J. Podolsk\'y\thanks{E--mail: {\tt podolsky@mbox.troja.mff.cuni.cz}}
\\
\\ Institute of Theoretical Physics, Charles University,\\
V Hole\v{s}ovi\v{c}k\'ach 2, 18000 Prague 8, Czech Republic.\\ }

\date{\today}

\maketitle

\begin{abstract}
A class of radiative solutions  of Einstein's field equations with a negative
cosmological constant and a pure radiation is investigated.
The space-times, which generalize the Defrise solution, represent exact
gravitational waves which interact with null matter and propagate in the
anti--de~Sitter  universe.  Interestingly, these solutions have homogeneous
and non-singular wave-fronts for all freely moving observers. We also study
properties of  sandwich and impulsive waves which can be constructed in this
class of space-times.
\end{abstract}



\section {Introduction}

There has been a growing interest in radiative  space-times which are not
asymptotically flat. In the last two decades, new exact solutions of this type,
representing gravitational waves in cosmology were
found and analyzed, for example in \cite{Gowdy}-\cite{Berger}  and elsewhere
(review of the  works can be found in \cite{CCM}, \cite{Verd}-\cite{JB2000}).
Some of these solutions can be interpreted as spatially inhomogeneous
cosmological models in which the homogeneity of the universe is broken due to
the presence of gravitational waves.

In this work we concentrate on presenting a physical interpretation of one
special class of exact type {\it N} solutions with a negative cosmological
constant $\Lambda$. These are generalizations of the Defrise solution
\cite{Defr} such that the profile of the wave may be arbitrary. Therefore,
particular gravitational waves propagating in an everywhere curved
anti--de Sitter universe can be constructed.

A unique feature of all such  space-times is that (for any bounded  wave-profile)
they are  non-singular, just like the well-known  homogeneous {\it pp}-waves
(plane waves) in a flat Minkowski background. This is demonstrated in the next section.
In Section~3 we  investigate  geodesics, and in Section~4 the geodesic deviation in
the generalized Defrise solutions. Remarks on the global structure  are presented
in Section~5. Using these solutions we construct, in Section~6, non-singular
sandwich waves in the anti--de Sitter universe and investigate their impulsive limits.
In particular,  geodesic motion in the Defrise sandwich and impulsive waves is
studied in Section~7.

\section {Generalized Defrise space-times}

The Defrise solution \cite{Defr}, \cite{KSMH} belongs to the Kundt class of non-twisting,
non-expanding  and shear-free space-times of  type {\it N}, in particular, to its
interesting  subclass which was found by Siklos \cite{Siklos}. For this subclass the
quadruple  Debever-Penrose null vector field {\bf k} is simultaneously the Killing
vector field. The metric can  be written in the simple form
\begin{equation}
\d s^2={\beta^2\over x^2}\, (\d x^2 + \d y^2 +2\,\d u\, \d v+H\,\d u^2)\ ,
\label{E2.1}
\end{equation}
where $\beta=\sqrt{-3/\Lambda}$, $\Lambda$ is a negative cosmological constant,
$x$ and $y$ are spatial coordinates, $v$ is an affine parameter along  rays
generated by ${\bf k}=\partial_v$, and $u$ is a retarded time. The function
$H(x,y,u)$ for the Defrise solution has the form $H=-x^{-2}$. It admits six
Killing vectors.

In the present paper we investigate solutions which generalize the Defrise solution
in such a way that the profile of the gravitational  wave may be arbitrary:
 we consider solution which can be written as
\begin{equation}
H={d(u)\over x^2}\ ,
\label{E2.3}
\end{equation}
where $d(u)$ is an arbitrary (bounded) function of $u$. The metric (\ref{E2.1}),
(\ref{E2.3})  satisfies the Einstein field equations with a negative cosmological
constant and a pure radiation field, $T_{\mu\nu}=\Phi k_\mu k_\nu$, where
$\Phi=-5\,d(u)/8\pi\beta^4$. This corresponds to null matter
propagating along the principal null congruence. For non-constant $d(u)$,
the space-time (\ref{E2.1}), (\ref{E2.3})  admits three Killing vectors:
 \  $\partial_v$,\  $\partial_y$,\ and\ $y\partial_v-u\partial_y$.

An interesting property of the solutions given by (\ref{E2.3}) is that these are
{\it non-singular} in the following sense. By inspecting the  components
of the curvature tensor with respect to orthonormal frames parallelly propagated along
any timelike geodesic (see \cite{Pod98}), it can be observed that all non-vanishing
components are proportional to one of the following functions:\
${\A}_+=-{C^2\over2}\,x^5\left({H_{,x}/x}\right)_{,x}$,\
${\A}_\times={C^2\over2}\,x^5\left({H_{,x}/ x} \right)_{,y}$,\   or\
$\M ={C^2\over2}\,x^3(xH_{,yy}-H_{,x})$,\
where $C$ is a constant (plus, in some cases, also to a constant term $\pm\Lambda/3$).
However, for the particular choice (\ref{E2.3}) of the function $H$ we obtain
\begin{equation}
{\A}_+=-4C^2\, d(u)\ ,
\qquad \M =C^2\, d(u) \ ,\qquad {\A}_\times=0 \ .
\label{E2.2}
\end{equation}
Therefore, for an arbitrary  bounded profile function $d(u)$, the frame components
of the {\it curvature tensor remains finite}, as seen by {\it any} timelike observer.
In this sense, all the  wave-surfaces $u=const$ are singularity-free. This is analogous
to a similar property for the well-known plane waves
in Minkowski space. Thus we may characterize the solution (\ref{E2.1}), (\ref{E2.3})
as representing exact waves in the anti--de Sitter universe with non-singular and
homogeneous wave-fronts. (In fact, it can easily be shown that the choice
(\ref{E2.3}) is the only non-trivial possibility for which the functions
${\A}_+$, $\M$ and ${\A}_\times$ are independent of the spatial coordinates
$x$ and $y$.)

The above solutions reduce to the anti--de Sitter solution in the regions where
$d(u)=0$. Therefore, we can easily construct space-times representing sandwich waves
 in the anti--de Sitter universe by considering $d(u)$ non-zero on a finite interval of
$u$ only, just as for standard sandwich gravitational waves
in flat space \cite{BPR}-\cite{BP}. Some properties  of these space-times will be
discussed  later in  Sections~6 and~7, together with the possibility
of constructing impulsive waves in the anti--de Sitter universe by
letting the profiles $d(u)$ approach the Dirac delta function.

\section {Geodesics}

The geodesic equations for the metric (\ref{E2.1}), (\ref{E2.3}) are
\begin{eqnarray}
&&\ddot v = 2\dot v {\dot x\over x} + 2C\,d(u) {\dot x\over x}
             -{\textstyle {1\over2}}C^2d'(u)x^2\ ,     \nonumber\\
&&{\ddot x\over x}-\left({\dot x\over x}\right)^2+ B^2x^2 + 2C\dot v
 +2C^2d(u)=0\ , \label{E2.8}\\
&&\dot y = Bx^2      \ ,\qquad
  \dot u = Cx^2\ , \nonumber\\
&&\left({\dot x\over x}\right)^2 + B^2x^2 + 2C\dot v + C^2d(u)
     -{\epsilon\over {\beta^2}}=0\ , \nonumber
\end{eqnarray}
where the dot denotes the derivative with respect to the affine parameter $\tau$,
the prime denotes the derivative with respect to $u$, and $B, C$ are constants.
Here the equations for $\dot y$ and $\dot u$ have already been integrated (this
can easily be achieved due to the existence of  corresponding Killing vectors). The last
equation in (\ref{E2.8}) is the normalization condition of the four-velocity, $u_\alpha u^\alpha = \epsilon$,
 where $\epsilon=-1,0$, or $+1$, for timelike, null or spacelike geodesics, respectively.

We may eliminate the terms containing the constant $B$ and the variable $\dot v$ by
subtracting the last equation in (\ref{E2.8}) from the second one. It is also
convenient to introduce a new variable $\xi=1/ x$,
with which we obtain the equation
\begin{equation}
\ddot\xi=\left(\, C^2d(u) + {\epsilon\over {\beta^2}}\,  \right)\,\xi\ .
\label{E2.11}
\end{equation}
For the particular case when the  profile function $d$ is constant, this is a simple
 equation for $\xi$. After obtaining $x(\tau)=\xi^{-1}(\tau)$, we
can then  integrate the  equations for the remaining functions $y(\tau)$, $u(\tau)$
and $v(\tau)$, see (\ref{simpl})-(\ref{geod++}). However, if $d(u)$ is a non-constant
function, then the equation (\ref{E2.11}) has to be solved simultaneously with the equation
\begin{equation}
\dot u={C\over \xi^2}\ .
\label{E2.12}
\end{equation}

Let us now present some particular solutions explicitly. The simplest geodesics arise when
$\dot x=0$. In this case (\ref{E2.11}) implies $C^2d(u)=-\epsilon/\beta^2$. When $C=0$,
we obtain a privileged class of null geodesics $v=A\tau+v_0$, $x=x_0$, $y=y_0$, $u=u_0$,
where $A, x_0, y_0, u_0, v_0$ are constants. These generate the null wave-fronts
$u=const$ in any generalized Defrise solution. If \ $C$\ is non-vanishing then
$d(u)$ must necessarily be a constant function, $d(u)=D=const$. The corresponding
geodesics are
\begin{equation}
v(\tau)=v_0+A\tau\ ,\quad x(\tau)=x_0\ ,\quad
y(\tau)=y_0+Bx_0^2\tau\ ,\quad u(\tau)=u_0+Cx_0^2\tau\ ,
\label{simpl}
\end{equation}
in which the constants have to satisfy the condition $B^2x_0^2+2AC=2\epsilon/\beta^2$.
When $D=0$ these are null geodesics in the anti--de~Sitter space-time. For $D>0$ the
geodesics (\ref{simpl}) are timelike and for $D<0$ spacelike. In these last two cases,
an additional restriction $C^2=1/(\beta^2|D|)$ applies. Note that the metric
(\ref{E2.1}), (\ref{E2.3}) for $d(u)=D\not=0$ represents the Defrise solution since
an arbitrary non-vanishing constant $D$ can be scaled to $\pm1$ by $u\to u/\sqrt{|D|}$,
$v\to\sqrt{|D|}\,v$.

In the general case $\dot x\not=0$ we have to solve (\ref{E2.11}) and (\ref{E2.12}).
For the anti--de~Sitter solution $(d(u)=D=0)$ and for the Defrise solution
$(d(u)=D=const\not=0)$, the system decouples. The equation for $\xi$ can immediately
be solved, yielding the following geodesics:
 \begin{equation}
\left.
\matrix{
{\displaystyle  v(\tau)=v_0+{\epsilon\,\tau\over\beta^2C}+
  {1+B^2x_0^2\over 2C(\tau-\tau_0)}\ , \qquad
    x(\tau)={x_0\over\tau-\tau_0}\ ,}\cr
\noalign{\medskip}
{\displaystyle  y(\tau)=y_0-{B\,x_0^2\over\tau-\tau_0}\ ,\quad
    u(\tau)=u_0-{C\,x_0^2\over\tau-\tau_0}\ , }}
\right\}
\qquad \hbox{for}\quad C^2D+{\epsilon\over\beta^2}=0\ ,
\label{geod0}
 \end{equation}

\begin{equation}
\left.
\matrix{
{\displaystyle
v(\tau)=v_0- CD\,\tau+
  {Ax_0^2\over a}\tan(a\tau-\tau_0)\ , \qquad
    x(\tau)={x_0\over\cos(a\tau-\tau_0)}\ ,}\cr
\noalign{\medskip}
{\displaystyle  y(\tau)=y_0+{B\,x_0^2\over a}\tan(a\tau-\tau_0)\ ,\quad
     u(\tau)=u_0+{C\,x_0^2\over a}\tan(a\tau-\tau_0)\ , }\cr
\noalign{\medskip}
\hskip20mm\hbox{with}\quad(B^2+2AC)x_0^2=C^2+D\epsilon\beta^{-2}\ ,}
\right\}
\qquad \hbox{for}\quad C^2D+{\epsilon\over\beta^2}<0\ ,
\label{geod-}
 \end{equation}

\begin{equation}
\left.
\matrix{
{\displaystyle
v(\tau)=v_0- CD\,\tau+
  {Ax_0^2\over a}\tanh(a\tau)\ , \qquad
    x(\tau)={x_0\over\cosh(a\tau)}\ ,}\cr
\noalign{\medskip}
{\displaystyle  y(\tau)=y_0+{B\,x_0^2\over a}\tanh(a\tau)\ ,\quad
     u(\tau)=u_0+{C\,x_0^2\over a}\tanh(a\tau)\ , }\cr
\noalign{\medskip}
\hskip20mm\hbox{with}\quad(B^2+2AC)x_0^2=C^2+D\epsilon\beta^{-2}\ ,}
\right\}
\qquad \hbox{for}\quad C^2D+{\epsilon\over\beta^2}>0\ ,
\label{geod+}
 \end{equation}

or
\begin{equation}
\left.
\matrix{
{\displaystyle
v(\tau)=v_0- CD\,\tau -
  {A\over a x_1}[\,x_0+x_1\tanh(a\tau)\,]^{-1}\ ,}\cr
\noalign{\medskip}
{\displaystyle
    x(\tau)=[\,x_0\cosh(a\tau)+x_1\sinh(a\tau)\,]^{-1}\ ,\hskip14mm}\cr
\noalign{\medskip}
{\displaystyle
y(\tau)=y_0-{B\over a x_1}[\,x_0+x_1\tanh(a\tau)\,]^{-1}\ ,\hskip14mm}\cr
\noalign{\medskip}
{\displaystyle
u(\tau)=u_0-{C\over a x_1}[\,x_0+x_1\tanh(a\tau)\,]^{-1}\ ,\hskip14mm}\cr
\noalign{\medskip}
\hskip3mm\hbox{with}\quad x_1\not=0\ ,\qquad
B^2+2AC=(x_0^2-x_1^2)(C^2+D\epsilon\beta^{-2})\ ,}
\right\}
\qquad \hbox{for}\quad C^2D+{\epsilon\over\beta^2}>0\ ,
\label{geod++}
 \end{equation}

\noindent
where $a=\sqrt{|C^2D+\epsilon\beta^{-2}|}$, and $\tau_0$, $x_1$ are arbitrary constants.

For a  non-constant profile $d(u)$ the geodesics can  be obtained by solving
simultaneously the equations (\ref{E2.11}) and (\ref{E2.12}) numerically, and
integrating subsequently $v(\tau)$ and $y(\tau)$. However, one important general
observation can be made for an arbitrary  solution. It is obvious from the equation
(\ref{E2.12}) that for {\it any}  geodesic such that\  $x(\tau)\to0$, one obtains\
$u(\tau)\to u_0$. Therefore, for a solution with an arbitrary wave-profile,
it follows that $d(u)\to d (u_0)=const=D$. This means that all geodesics behave
asymptotically according to one of the corresponding possibilities described
by (\ref{geod0}), (\ref{geod+}), or (\ref{geod++}), as $x(\tau)\to0$.

\section {Geodesic deviation}

It has been shown previously in \cite{Pod98} that the equation of geodesic deviation
along any timelike geodesic, given by (\ref{E2.8}), in a suitably chosen
orthonormal frame $\{ {\bf e}_{a'} \}$
\begin{eqnarray}
e^\mu_{(0)}&=&u^\mu=\left(\dot v,\dot x,Bx^2,Cx^2\right)\ , \nonumber\\
e^\mu_{(1')}&=&\left(-{1\over{\beta C}}{\dot x\over x} \;,{x\over\beta},0,0\right)\ , \nonumber\\
e^\mu_{(2)}&=&{x\over\beta}\left(-{B\over C}\;,0,1,0\right)\ ,\label{E2.14}\\
e^\mu_{(3')}&=&\left(\dot v+{1\over{\beta^2C}}\;,\dot x, Bx^2,Cx^2\right)\ ,
\nonumber
\end{eqnarray}
can be written as
\begin{eqnarray}
\ddot Z^{(1')} &=&{\frac{\Lambda}{3}}Z^{(1')}\ -\  \A _+ Z^{(1')}
\ , \nonumber\\
\ddot Z^{(2)}\,&=&{\frac{\Lambda}{3}}Z^{(2)}\ \>+\ \M\>  Z^{(2)}\>\ , \label{devi}\\
\ddot Z^{(3')} &=&{\frac{\Lambda}{3}}Z^{(3')}\ . \nonumber
\end{eqnarray}
The amplitudes ${\A}_+$ and  $\M$  are ${\A}_+=-4C^2\, d(u)$ and
$\M =C^2\,d(u)$ (see (\ref{E2.2})), $Z^{(i)} = e^{(i)}_\mu Z^\mu$ denote frame
components of the displacement vector connecting two neighbouring free test
particles, and $\ddot Z^{(i)} = e^{(i)}_\mu {{D^2Z^\mu}\over{d\tau^2}}$ are
their relative accelerations.

Equations (\ref{devi}) suggest the following physical interpretation of the generalized
Defrise space-times. In the regions where $d(u)=0$ the functions ${\A}_+$ and  $\M$
vanish. The solution reduces to the anti--de~Sitter space-time in which
all test particles move isotropically one with respect to the other,
$\ddot Z^{(i)}={\Lambda\over3}Z^{(i)}$. Thus, the terms proportional to $\Lambda$ in
(\ref{devi}) represent the influence of the {\it anti--de~Sitter background}. If the
amplitudes ${\A}_+$ and  $\M$ do not vanish (which is for $d(u)\not=0$), these
background motions of  particles are influenced also by the {\it effect
of the gravitational wave} combined with that of the {\it null matter}. Both the
gravitational wave and  the pure radiation propagate in the spacelike direction of
${\bf e}_{(3')}$ and have a {\it transverse} character since only motions in
the perpendicular directions  ${\bf e}_{(1')}$ and ${\bf e}_{(2)}$ are affected.

Note however that the direction of propagation ${\bf e}_{(3')}$ in {\it not}
parallelly transported. Instead, it uniformly rotates with angular velocity
given by\ $1/\beta=\sqrt{-\Lambda/3}$ with respect to frames $\{{\bf e}_{(a)}\}$
parallelly propagated along any timelike geodesic,
\begin{equation}
{\bf e}_{(1')}=\cos(\tau/\beta)\,{\bf e}_{(1)} -
    \sin(\tau/\beta)\,{\bf e}_{(3)}\ ,\quad
{\bf e}_{(3')}=\sin(\tau/\beta)\,{\bf e}_{(1)} +
    \cos(\tau/\beta)\,{\bf e}_{(3)}\ .
    \label{rot}
\end{equation}
This effect has been demonstrated for all solutions of the Siklos class in
\cite{Pod98}.

It is well-known that the effect of pure  vacuum gravitational waves with the `+'
polarization mode on relative motions of the  test particles can
be described by the equations (\ref{devi}) with    $\M={\A}_+$
(see e.g. \cite{BiPo}). However, in our case $\M\not={\A}_+$. Nevertheless, we
may interpret the effect by introducing a decomposition $\M={\A}_+ +{\cal P}$,
where ${\cal P}=5C^2d(u)$ . Substituting  for $\M$ in (\ref{devi}) we observe
that the influence on particles given by the anti--de~Sitter background and a
`pure' gravitational wave with the amplitude $\A_+$, superpose with the effect
given by the term ${\cal P}\,Z^{(2)}$. This is responsible for an additional
acceleration in the direction of  ${\bf e}_{(2)}$  due to the presence of null
matter.

As in the case of vacuum Siklos space-times, we can rewrite the
equation of geodesic deviation in a form that is suitable for integration
(note that $\ddot Z^{(i)}$ does not represent the total time derivative of
$Z^{(i)}(\tau)$ for $i=1',3'$ since ${\bf e}_{(1')},{\bf e}_{(3')}$ are  not
parallelly transported). Using the relations (23) given in \cite{Pod98},
the  system (\ref{devi}) can be written as
\begin{eqnarray}
{{d^2Z^{(1')}}\over{d\tau^2}}+4\left[{{1\over{\beta^2}}-C^2\,d(u(\tau))}\right]
    Z^{(1')} &=& -{2\over\beta}C_1\ , \nonumber\\
{{d^2Z^{(2)}}\over{d\tau^2}}\;+\left[{{1\over{\beta^2}}-C^2\,d(u(\tau))}\right]
    Z^{(2)}\>&=& 0\ , \label{E3.28}\\
Z^{(3')}&=&\int \left({2\over\beta}Z^{(1')}+C_1\right)\, d\tau\ ,
\nonumber
\end{eqnarray}
where $C_1$ is a constant. These decoupled equations can be integrated provided
the geodesic function $u(\tau)$ is known. However, there  exists a  solution,
along {\it any}  geodesic in {\it any} generalized Defrise solution, given by
$Z^{(1')}=0=Z^{(2)}$, $Z^{(3')}=Z_0=const$,  i.e. using (\ref{rot}),
\begin{equation}
Z^{(1)}\;=Z_0\,\sin(\tau/\beta),\quad Z^{(2)}=0,\quad Z^{(3)}\;=Z_0\,\cos(\tau/\beta)\ .
\label{rotac}
\end{equation}
The  particles may  corotate uniformly in {\it circles} with constant
angular velocity $\sqrt{-{\Lambda}/{3}}$.

Note also that the equations (\ref{E3.28}) are {\it independent} of $x(\tau)$, $y(\tau)$
and $v(\tau)$ which again demonstrates the homogeneity of the wave-fronts
$u=const$. Thus all timelike observers on a given $u$ will view the  same
relative motions of the surrounding test particles.

Let us finally present the complete solution of (\ref{E3.28}) for the case when
$d(u)=D=const$\ :

 \begin{equation}
\left.
\matrix{
{\displaystyle  Z^{(1')}(\tau)=-{C_1\over\beta}\,\tau^2+C_2\,\tau+C_3\ ,}\hskip35mm\cr
\noalign{\medskip}
{\displaystyle  Z^{(2)}(\tau)\>=\ C_4\,\tau+C_5\ , }\hskip50mm\cr
\noalign{\medskip}
{\displaystyle  Z^{(3')}(\tau)=-{2C_1\over3\beta^2}\,\tau^3+{C_2\over\beta}\,\tau^2
  +\left({2C_3\over\beta}+C_1\right)\,\tau+C_6\ ,}
}
\right\}
\quad \hbox{for}\quad C^2D-{1\over\beta^2}=0\ ,
\label{exgeod0}
 \end{equation}

 \begin{equation}
\left.
\matrix{
{\displaystyle  Z^{(1')}(\tau)=-{C_1\over2a^2\beta}+C_2\cos(2a\tau)+C_3\sin(2a\tau)\ ,}\hskip18mm\cr
\noalign{\medskip}
{\displaystyle  Z^{(2)}(\tau)\>=\ C_4\cos(a\tau)+C_5\sin(a\tau)\ , }\hskip37mm\cr
\noalign{\medskip}
{\displaystyle  Z^{(3')}(\tau)=-{C_1C^2D\over a^2}\,\tau+{C_2\over a\beta}\sin(2a\tau)
   -{C_3\over a\beta}\cos(2a\tau)+C_6\ ,}
}
\right\}
\quad \hbox{for}\quad C^2D-{1\over\beta^2}<0\ ,
\label{exgeod-}
 \end{equation}

 \begin{equation}
\left.
\matrix{
{\displaystyle  Z^{(1')}(\tau)={C_1\over2a^2\beta}+C_2\cosh(2a\tau)+C_3\sinh(2a\tau)\ ,}\hskip18mm\cr
\noalign{\medskip}
{\displaystyle  Z^{(2)}(\tau)\>=\ C_4\cosh(a\tau)+C_5\sinh(a\tau)\ , }\hskip34mm\cr
\noalign{\medskip}
{\displaystyle  Z^{(3')}(\tau)={C_1C^2D\over a^2}\,\tau+{C_2\over a\beta}\sinh(2a\tau)
   +{C_3\over a\beta}\cosh(2a\tau)+C_6\ ,}
}
\right\}
\quad \hbox{for}\quad C^2D-{1\over\beta^2}>0\ ,
\label{exgeod+}
 \end{equation}
where $C_i$ are constants and $a$ is again given by $a=\sqrt{|C^2D-\beta^{-2}|}$. These
relations describe all possible relative motions of nearby particles in the
anti--de~Sitter and the Defrise space-times.

In particular, for the anti--de~Sitter universe, $D=0$, so that
only the motions given by (\ref{exgeod-}) are allowed. Using the relation
(\ref{rot}), these can be written in a parallelly propagated frame as
$Z^{(i)}(\tau)=A_i\cos(\tau/\beta+\delta_i)$,
where $A_i$ and $\delta_i$, $i=1,2,3$, are constants.

On the other hand, we may consider the limit $\Lambda\to 0$, i.e. $1/\beta\to 0$,
in which case the rotation of the frame (\ref{rot}) vanishes,
$Z^{(1')}\to Z^{(1)}$, $Z^{(3')}\to Z^{(3)}$. Assuming $D=-\omega^2$, where
$\omega$ is some positive constant, we get $a\to |C|\omega$, and equations
(\ref{exgeod-}) become
\begin{eqnarray}
Z^{(1)}&\approx& A_1\cos(\,2|C|\omega\,\tau+\delta_1\,)\ , \nonumber\\
Z^{(2)}&=     & A_2\cos(\,\>|C|\omega\,\tau+\delta_2\,)\ , \label{aprox}\\
Z^{(3)}&\approx& A_3\,\tau+\delta_3\ . \nonumber
\end{eqnarray}
The particles move freely --- as in Minkowski space --- along the direction
${\bf e}_{(3)}$ which is the direction of propagation of the waves. In the transverse
plane the relative motions of nearby test particles follow the famous closed
Lissajous figures.

\section {On the global structure}

The metric (\ref{E2.1}), (\ref{E2.3})  indicates that the space-times are regular
everywhere except possibly at $x=0$ and/or $x=\infty$. We shall investigate these
regions in detail. Let us perform the transformation
\begin{eqnarray}
 && \eta\ =-{\beta\cos(T/\beta)}/{\cal D}\ , \quad
  x   \ =\ {\beta\cos\chi}/{\cal D}     \ , \label{E6.16} \\
 && y \ =\ {\beta\sin\chi\cos\vartheta}/{\cal D}  \ ,\quad
  z   \ =\ {\beta\sin\chi\sin\vartheta\cos\varphi}/{\cal D} \ , \nonumber
\end{eqnarray}
where $\eta=(u-v)/\sqrt2$, $z=(u+v)/\sqrt2$, and
${\cal D}=\sin(T/\beta)+\sin\chi\sin\vartheta\sin\varphi$. This puts the
 metric of generalized Defrise space-times into the form
\begin{eqnarray}
\d s^2&=&\frac{\beta^2}{\cos^2 \chi}
  \left\{-\frac{\d T^2}{\beta^2}+\d\chi^2+\sin^2\chi(\d\vartheta^2+
       \sin^2\vartheta\, \d\varphi^2)\right\} \label{E6.17} \\
&+&\frac{d(u(T,\chi,\vartheta,\varphi))}{2\cos^4\chi}
  \Bigg\{
  [1-\cos(T/\beta+\varphi)\sin\chi\sin\vartheta]\frac{\d T}{\beta}
  +\sin(T/\beta+\varphi)\cos\chi\sin\vartheta\,\d\chi \nonumber\\
&&\qquad +\sin(T/\beta+\varphi)\sin\chi\cos\vartheta\, \d\vartheta
+\sin\chi\sin\vartheta[\cos(T/\beta+\varphi)-\sin\chi\sin\vartheta]
  \,\d\varphi \Bigg \}^2\ , \nonumber
\end{eqnarray}
where the argument of the profile function $d$ is
\begin{equation}
u(T,\chi,\vartheta,\varphi)=(\beta/\sqrt2)
{-\cos(T/\beta)+\sin\chi\sin\vartheta\cos\varphi\over
\sin(T/\beta)+\sin\chi\sin\vartheta\sin\varphi}\ .
\label{arg}
\end{equation}
For $d\equiv 0$, this is the well-known form of the anti--de~Sitter solution in
 global coordinates (cf. \S 5.2 in \cite{HE} where $\cosh r=1/\cos\chi$) which
is used in the literature to construct the Penrose diagram. Choosing
the conformal factor $\Omega=\beta^{-1}\cos\chi$, the boundary $\Omega=0$ of the
anti--de Sitter manifold   (corresponding to $\chi=\pi/2$) represents null
and spacelike infinity which  can be thought of as a  timelike surface with
 topology $R\times S^2$.

The metric form (\ref{E6.17}) demonstrates explicitly that, for bounded profiles
$d(u)$, the space-times are regular everywhere, except at $\chi=\pi/2$.
Therefore, $x=\infty$ (which corresponds to  ${\cal D}=0$, $\chi\not=\pi/2$)
is only a {\it coordinate} singularity. In fact, by inspecting the particular
geodesics (\ref{geod0}), (\ref{geod-}), (\ref{geod++}) it can be seen that $x=\infty$
is reached  at {\it finite} values of the affine parameters. This indicates
that  $x=\infty$ is not a boundary of the manifold, which can thus be extended
beyond ${\cal D}=0$. This continuation is achieved by putting the solution into
the form (\ref{E6.17}) and considering the full range of the coordinates,
$T\in(-\infty, +\infty)$, $\chi\in[0,\pi/2)$, $\vartheta\in[0,\pi]$, $\varphi\in[0,2\pi)$.
(Let us also remark that even with the help of the coordinate $\xi=1/x$, the
geodesics (\ref{geod0})-(\ref{geod++}) can analytically be extended through
$x=\infty$ which corresponds to $\xi=0$.)

We now investigate the singularity at $x=0$. This is mapped to $\chi=\pi/2$,
i.e. it is located at the ``anti--de~Sitter-like'' infinity given by the boundary
$\Omega=0$. We have already emphasized (see end of  Section 3) that {\it all}
geodesics approaching $x=0$ behave asymptotically according to (\ref{geod0}),
(\ref{geod+}), or (\ref{geod++}). Therefore, an {\it infinite} value of the
affine parameter $\tau$ is  required  to reach $x=0$. This supports our observation
that this singularity is located at the very boundary of the manifold. Moreover,
all components of the curvature tensor in the orthonormal frame parallelly
propagated along timelike geodesics are given by (\ref{E2.2}), and obviously
{\it remain finite} even as $x\to 0$. This indicates that the singularity at $x=0$
is quasiregular (according to  the classification scheme introduced in \cite{ES}),
i.e. it has a ``topological'' rather than a ``curvature'' character.

Finally, let us transform the generalized Defrise solutions (\ref{E2.1}),
(\ref{E2.3}) using
\begin{equation}
2x={\pm1\over \cosh\theta+\sinh\theta\cos\phi}\ ,\qquad\qquad
2y={\sinh\theta\sin\phi\over \cosh\theta+\sinh\theta\cos\phi}\ ,
\label{hyp}
\end{equation}
to obtain
\begin{eqnarray}
&&\d s^2=\beta^2\, (\d \theta^2 + \sinh\theta\,\d \phi^2)
 +8\beta^2\,(\cosh\theta+\sinh\theta\cos\phi)^2\,\d u\, \d v \nonumber\\
&&\qquad\qquad +16\beta^2\,(\cosh\theta+\sinh\theta\cos\phi)^4\,\,d(u)\,\d u^2\ ,
\label{hyper}
\end{eqnarray}
where $\theta\in[0,\infty)$, $\phi\in[0,2\pi)$, $u,v\in(-\infty,+\infty)$.
The singularity at  $x=0$ is now given by $\theta=\infty$. The form
(\ref{hyper}) of the solutions exhibits explicitly the geometry of the
wave-surfaces $u=const$\ : these are two-dimensional {\it hyperboloidal
surfaces} of constant negative curvature $-\beta$.

\section{Sandwich and impulsive waves in the anti--de~Sitter universe}

Using the above results, we may now consider the construction of sandwich
(gravitational plus null matter) waves in the anti--de~Sitter space. Obviously,
these are described by the metric (\ref{E2.1}) , (\ref{E2.3}), or equivalently
by (\ref{hyper}), if the wave-profile function $d(u)$ is non-vanishing on a
{\it finite}\ interval, say $u\in[u_1,u_2]$, only. In such a case, the sandwich
wave has a finite duration and extends between two hyperboloidal surfaces
$u_1$ and $u_2$ representing the front and the end of the wave. In front of
the propagating sandwich wave of type {\it N}\ , for $u<u_1$, and also behind
it, for $u>u_2$, there are two anti--de~Sitter regions which are conformally flat
and maximally symmetric. The situation is analogous to the well-known case
in Minkowski universe in which, however, the plane waves propagate through
the flat space \cite{BPR}-\cite{BP}.

To obtain a better understanding of the geometry of these sandwich waves let
us recall that the anti--de~Sitter universe can be seen as a four-dimensional
hyperboloid\ $-Z_0^2+Z_1^2+Z_2^2+Z_3^2-Z_4^2=-\beta^2$, embedded in a
five-dimensional flat space-time \ $\d s_0^2=-\d Z_0^2+\d Z_1^2+\d Z_2^2+\d Z_3^2-\d Z_4^2$,
with two time coordinates $Z_0$ and $Z_4$ (see e.g. \cite{HE}). This is shown in
Fig. 1. The most natural global parametrization is
\begin{eqnarray}
 &&Z_0\ =-\beta\,\cos(T/\beta)/\cos\chi\ , \qquad
   Z_4\ =\ \beta\,\sin(T/\beta)/\cos\chi\ , \nonumber \\
 &&Z_1\ =\ \beta\,\tan\chi\sin\vartheta\cos\varphi\ ,\qquad\
   Z_2\ =\ \beta\,\tan\chi\cos\vartheta , \label{para}\\
 &&Z_3\ =\ \beta\,\tan\chi\sin\vartheta\sin\varphi\ ,\nonumber
\end{eqnarray}
which gives the coordinate system (\ref{E6.17}) for $d=0$. The beginning and the end
of the sandwich wave, given by $u=u_1$ and $u=u_2$, can now be visualized on the
above hyperboloid. (Note that when $d$ is small, the  waves can be considered
to represent a perturbation of the anti--de~Sitter hyperboloid so that also the
``inner'' wave-surfaces $u=const\in(u_1, u_2)$ can be treated similarly.) Using
(\ref{arg}) and (\ref{para}) we obtain
\begin{equation}
Z_0+Z_1=(\sqrt2/\beta)\, u\, (Z_3+Z_4)\ . \label{uconst}
\end{equation}
Each wave-front is thus located on the two-dimensional intersection of the
hyperboloid with the null hyperplane (\ref{uconst}) for a fixed $u$. In particular,
the wave-surface $u=0$ corresponds to $Z_0+Z_1=0$, which is a two-dimensional
hyperboloidal surface $Z_4^2-Z_2^2-Z_3^2=\beta^2$. The wave-fronts $u=\pm\infty$
are given by $Z_3+Z_4=0$, corresponding to $Z_0^2-Z_1^2-Z_2^2=\beta^2$. The intersections
for general $u$  given by (\ref{uconst}) are more difficult to visualize. In
Fig.~1 we draw them after supressing two space coordinates, $Z_2=0=Z_3$, and we
also assume that $u_1<0$ and $u_2>0$. Nevertheless, this has a disadvantage that
the null character of these intersections is not seen explicitly, except for the
wave-surface $u=0$. However, the picture still gives a useful insight into the
geometry of the sandwich waves in the anti--de~Sitter universe. We can also
easily observe that the complete picture should contain {\it two } sandwich waves,
first at $Z_4>0$ and another one at $Z_4<0$. Any observer moving around the
anti--de~Sitter hyperboloid in closed timelike loops would first observe a
sandwich wave propagating in one direction, then the second propagating in the
opposite direction, then again the first one, and so on in an endless cycle.
(Alternatively, this can be considered to represent only one wave which ``bounces''
back and forth from one side of the universe to the other.) These sandwich waves
are also shown in the conformal diagram in Fig.~2.

The  five-dimensional formalism also enables us  to construct the
Defrise-type {\it impulsive} waves in the anti--de~Sitter universe. By combining
the transformation (\ref{E6.16}) with the parametrization (\ref{para}), we may
write the  generalized Defrise solutions as
$\d s^2=\d s_0^2+(\beta^2/x^4)\,d(u)\,\d u^2$, where $\d s_0^2$ is the metric
on the anti--de~Sitter hyperboloid. Let us now consider a sequence of wave-profiles
$d(u)$ approaching the Dirac delta-distribution $\delta(u)$ localized  on
the null hypersurface $u=(\beta/\sqrt2)(Z_0+Z_1)/(Z_3+Z_4)=0$. Straightforward
calculation gives the distributional limit
\begin{eqnarray}
&&d s^2=-\d Z_0^2+\d Z_1^2+\d Z_2^2+\d Z_3^2-\d Z_4^2
\ +\ H\,\delta(Z_0+Z_1)\,(\d Z_0+\d Z_1)^2\ , \label{impuls}\\
&&\hskip20mm\hbox{where}\qquad \sqrt2\beta^5\,H=(Z_3+Z_4)^3\ , \nonumber
\end{eqnarray}
which describes the metric of the implusive gravitational wave plus the null-matter wave.
This is the  particular solution which belongs to a general class on non-expanding
impulsive waves in the anti--de~Sitter universe, presented in \cite{PG}, \cite{Pod98a}.
The geometry of the impulsive surfaces $Z_0+Z_1=0$ has been discussed in detail
in \cite{PG97}.

\section{Geodesics in sandwich and impulsive Defrise waves}

Finally, we present the simplest example of these  sandwich
waves  given by profile functions of the form
\begin{equation}
d(u)\ =\ D\>[\,\Theta(u)\>-\>\Theta(\,u-u_2\,)\,]\ , \label{sim}
\end{equation}
where $u_2$ is a positive constant and $\Theta$ is the Heaviside step function. Since
$d(u)=0$ for $u<0$ and for $u>u_2$, whereas $d(u)=D$ for $u\in[0,u_2]$, the
solutions given by (\ref{sim}) represent the Defrise sandwich waves with
{\it constant} amplitudes propagating in the anti--de~Sitter universe.

Using the explicit forms of the geodesics (\ref{simpl})-(\ref{geod++}) it is
possible to  find motion in these space-times. We concentrate here on a
privileged  class of timelike geodesics which in the (complete) anti--de~Sitter
spacetime are given by
\begin{eqnarray}
 &&Z_0\ =\>-\beta\,\cos(\tau/\beta)\ ,  \quad
   Z_1\ =\ \beta\,\sin(\tau/\beta)\,\sinh\psi\ , \quad
   Z_4\ =\ \beta\,\sin(\tau/\beta)\,\cosh\psi\ ,  \label{g}\\
&& Z_2\ =\ 0\ =\ Z_3\ , \nonumber
\end{eqnarray}
where $\tau$ is the proper time and $\psi\in(-\infty,+\infty)$ is an arbitrary
constant which parametrizes the specific geodesic from the above family.  Observers
following these geodesics move around the hyperboloid in closed timelike loops given
by  the intersections of the hyperboloid with the planes $Z_1=\tanh\psi\,Z_4$, as
indicated in Fig.~1. At $\tau=0$ all observers are located at one point, $Z_0=-\beta$,
$Z_1=0=Z_4$, and start moving with different velocities. At $\tau=(\pi/2)\,\beta$
they reach their maximum distance $Z_1$ from the ``space origin'' $Z_1=0$.
Subsequently, they converge back and all meet again simultaneously  at the point
$Z_0=\beta$, $Z_1=0=Z_4$ at  $\tau=\pi\,\beta$. Then continue  on the other side of the
anti--de~Sitter hyperboloid ($Z_4<0$) symmetrically, and return back to the starting
point. The existence of  these specific geodesics is caused by the presence
of a negative cosmological constant which has the effect of a universal
attractive force.

Our objective here is to investigate how the ``focusing'' effect described
above is changed when the observers pass through a sandwich wave of the
Defrise type. We assume  the wave-profile $d(u)$ has the form (\ref{sim})
so that there are three regions:\ I. The anti--de~Sitter region $u<0$ in
front of the wave,\ II. The Defrise wave for $0<u<u_2$\ , \ and\
III. Another anti--de~Sitter region $u>u_2$ behind the wave (see Fig.~3).

We start in the region I. with the privileged geodesics  (\ref{g}) which can be
written in the coordinates of the anti--de~Sitter metric (\ref{E2.1}) using the
corresponding parametrization
\begin{equation}
\eta={\beta Z_0\over Z_3+Z_4}\ ,\quad
x={\beta^2\over Z_3+Z_4}\ ,\quad
y={\beta Z_2\over Z_3+Z_4}\ ,\quad
z={\beta Z_1\over Z_3+Z_4}\ ,\label{par}
\end{equation}
(which follows from (\ref{E6.16}) and (\ref{para})) as
\begin{eqnarray}
&&x(\tau)={\beta\over\cosh\psi\,\sin(\tau/\beta)}\ ,\nonumber\\
&&y(\tau)=0\ ,\nonumber\\
&&u(\tau)={\beta\over\sqrt2}\,\tanh\psi
        \>-\>{\beta\over\sqrt2}\,{\cot(\tau/\beta)\over\cosh\psi}\ , \label{gI}\\
&& v(\tau)={\beta\over\sqrt2}\,\tanh\psi
        \>+\>{\beta\over\sqrt2}\,{\cot(\tau/\beta)\over\cosh\psi}\ . \nonumber
\end{eqnarray}
These can easily be identified in the general class of timelike geodesics
(\ref{geod-}). The geodesics (\ref{gI}) start at $\tau=0$ on the hypersurface
$u=-\infty$ and continue through the anti--de~Sitter region $u<0$ until they reach the
front $u=0$ of the sandwich wave, as indicated in Fig.~3. Different observers with
their specific values of the parameter $\psi$ reach the wave in different
times $\tau_f$ which are given by
\begin{equation}
\cot(\,\tau_f/\beta\,)=\sinh\psi\ .  \label{tf}
\end{equation}
This implies that observers with higher values of $\psi$ encounter  the wave sooner,
so that the wave propagates from right to left (from positive to negative
values of $Z_1$).

Now, we wish to extend the geodesics (\ref{gI}) into the sandwich-wave region II. We
assume that the geodesic functions $x(\tau)$, $y(\tau)$, $u(\tau)$ and $v(\tau)$ are
{\it continuous} across $u=0$, i.e. at $\tau=\tau_f$. For simplicity we consider here
only  geodesics for which $y(\tau)\equiv0$ at any $\tau$. In addition, we require that
$\dot x(\tau)$ is also a continuous function of the proper time. Note that $\dot u$
is  continuous as a consequence of the relation $\dot u=Cx^2$, provided the constant
$C$ has the same value in all the three regions. However, we {\it cannot} require
$\dot v$ to be continuous. In fact, by inspecting the geodesic equations (\ref{E2.8}), in
particular the last equation representing the normalization condition, it is
obvious that such an additional assumption would  be inconsistent with (\ref{sim}).
Instead, we have  to  prescribe a {\it discontinuity} in $\dot v$ at $u=0$ and
$u=u_2$ given by
\begin{eqnarray}
\dot v(u\to 0_+)&=&\dot v(u\to0_-)\,\,-\textstyle{{1\over2}} CD\ ,  \label{jumpI}\\
\dot v(u\to {u_2}_+)&=&\dot v(u\to {u_2}_-)+\textstyle{{1\over2}} CD\ .  \label{jumpII}
\end{eqnarray}

It is now straightforward to find among (\ref{geod-}) the explicit forms of the geodesics
in the Defrise wave-zone region II. given by $0\le u\le u_2$. These are
\begin{eqnarray}
&&x(\tau)=\sqrt{{1-(D/2\beta^2)\cosh^2\psi}\over{1-(D/2\beta^2)}}
    \,{\beta\over\cosh\psi\,\cos(a\tau-\tau_0)}\ ,\nonumber\\
&&u(\tau)={\beta\tanh\psi\over\sqrt2[1-(D/2\beta^2)]}\,
        \left[1+ {\sqrt{1-(D/2\beta^2)\cosh^2\psi}\over\sinh\psi}\,
        \tan(a\tau-\tau_0)\right]\ , \label{gII}\\
&& v(\tau)={D\over\sqrt2\beta^2}\cosh\psi\,(\tau_f-\tau)\ +\
       {\beta\over\sqrt2}\tanh\psi\, \left[1-
       {\sqrt{1-(D/2\beta^2)\cosh^2\psi}\over\sinh\psi}\,
       \tan(a\tau-\tau_0)\right]\ , \nonumber
\end{eqnarray}
in which the constant $\tau_0$ is given by
\begin{equation}
\tan(\,a\tau_f-\tau_0\,)=-{\sinh\psi\over \sqrt{1-(D/2\beta^2)\cosh^2\psi}}\ ,\label{t0}
\end{equation}
and $a=\sqrt{1-(D/2\beta^2)\cosh^2\psi}\,/\,\beta$. The observers with specific
$\psi$ move along the geodesics (\ref{gII}) until they reach the end of the Defrise
sandwich wave,  $u=u_2$,  at their  proper times $\tau_e$ given by
\begin{equation}
\tan(\,a\tau_e-\tau_0\,)={(\sqrt2/\beta)[1-(D/2\beta^2)]\,u_2\cosh\psi-\sinh\psi
\over \sqrt{1-(D/2\beta^2)\cosh^2\psi}}\ .\label{te}
\end{equation}

Using (\ref{jumpII}) we may similarly extend the geodesic across $u=u_2$ into the
anti--de~Sitter region III. behind the wave:
\begin{eqnarray}
&&x(\tau)={\beta\sqrt{K}\over\cosh\psi\,\cos(\tau/\beta-\tau_0^\ast)}\ ,\nonumber\\
&&u(\tau)=\Big(1-K[1-(D/2\beta^2)]\Big)\,u_2  +  {\beta K\over\sqrt2}\tanh\psi\,
    \Big[1+{\tan(\tau/\beta-\tau_0^\ast)\over\sinh\psi} \Big]\ , \label{gIII}\\
&& v(\tau)={D\over\sqrt2\beta^2}\cosh\psi\,(\tau_f-\tau_e)\ +\
       {\beta\over\sqrt2}\tanh\psi\, \left[1-
       {\tan(\tau/\beta-\tau_0^\ast)\over\sinh\psi}\, \right]\ , \nonumber
\end{eqnarray}
in which the constant $\tau_0^\ast$ is given by
\begin{equation}
\tan(\,\tau_e/\beta-\tau_0^\ast\,)=(\sqrt2/\beta)[1-(D/2\beta^2)]\,u_2\cosh\psi-\sinh\psi\ ,\label{t0a}
\end{equation}
and
\begin{equation}
K={{1+(2/\beta^2)[1-(D/2\beta^2)]\,u_2^2-(2\sqrt2/\beta)\,u_2\tanh\psi}
  \over{1+(2/\beta^2)[1-(D/2\beta^2)]^2\,u_2^2-(2\sqrt2/\beta)[1-(D/2\beta^2)]
    \,u_2\tanh\psi}}\ .  \label{K}
\end{equation}

It is obvious that the geodesics (\ref{gII}) and (\ref{gIII}) reduce to (\ref{gI})
when $D=0$ (implying $a=1/\beta$, $K=1$, and $\tau_0=\pi/2=\tau_0^\ast\,$), and also
for $u_2\to 0$ with finite $D$ (in which case $K\to1$, $\tau_e\to \tau_f$,
$\tau_0^\ast\to\pi/2$, and the Defrise wave-region II. disappears).

The above geodesics (\ref{gI}) in the anti--de~Sitter universe reconverge to the ``space''
origin $Z_1=0$ all at the same time $\tau=\pi\beta$. Indeed, using (\ref{par}) we
obtain $Z_1=\beta\,z/x=(\beta/\sqrt2)\,(u+v)/x$, but from (\ref{gI}) it follows that
$1/x$ is proportional to $\sin(\tau/\beta)$ which vanishes {\it independently} of $\psi$.
However, if the motion of the observers is influenced by the sandwich wave, it follows
from (\ref{gIII}) that $Z_1\sim1/x\sim\cos(\tau/\beta-\tau_0^\ast)$. Thus, the observers
return back to $Z_1=0$ at times $\tau=(\tau_0^\ast+\pi/2)\beta$. However, these  are
now generally different and individual since the parameter $\tau_0^\ast$,
given by (\ref{t0a}), (\ref{te}),  (\ref{t0}) and (\ref{tf}),
is a  complicated function of $D$, $u_2$ and, in particular, of $\psi$.

Let us finally  consider the geodesic motion in the Defrise-type {\it impulsive}
wave (\ref{impuls}) in the anti--de~Sitter universe. To this end we assume
a sequence of sandwich gravitational plus null matter waves given by
(\ref{sim}), in which the parameters $D$ and $u_2$ satisfy the normalization
condition \ $D\,u_2=-1$\ (we require $D<0$, but this is the physically
interesting case for which the amplitude  $\Phi=-5\,D/8\pi\beta^4$ in the
pure radiation energy-momentum tensor $T_{\mu\nu}=\Phi k_\mu k_\nu$ is
positive). The geodesics in the impulsive Defrise wave  can now  be obtained
from geodesics  in the corresponding  sandwich waves (\ref{gI})-(\ref{K})
by assuming the limit $u_2\to0$, i.e. $D\to-\infty$, for which the sequence
of $d(u)$ approaches the Dirac distribution $\delta(u)$. Straightforward
calculations using (\ref{t0}) and (\ref{te}) yield
\begin{equation}
\tau_e-\tau_f={1\over a}\arctan\left(
{(\sqrt2/\beta)\,u_2\,\sqrt{1-(D/2\beta^2)\cosh^2\psi}\over
\cosh\psi-(\sqrt2/\beta)\,u_2\sinh\psi}\right)\sim{\sqrt2\, u_2\over\cosh\psi}\to 0\ ,
\label{tetf}
\end{equation}
so that $\tau_e\to \tau_f$. This is  expected since the sandwich wave region II.
vanishes in the limit $u_2\to0$. The geodesics in the anti--de~Sitter region III. behind
the impulse ($u>0$ in the limit) are given by (\ref{gIII}). In particular,
we can evaluate the  functions at the time $\tau_e$  at which the specific observers
stop interacting with the impulsive wave localized at $u=0$,
\begin{equation}
x(\tau_e)=\beta\ ,\qquad
y(\tau_e)=0\ ,\qquad
u(\tau_e)=0\ ,\qquad
v(\tau_e)=\sqrt2\beta\tanh\psi\ +\ {1\over 2\beta^2}\ .\label{jIII}
\end{equation}
This can now be compared with the corresponding values obtained from the geodesics
(\ref{gI}) in the region I. in front of the wave ($u<0$) at $\tau=\tau_f$,
\begin{equation}
x(\tau_f)=\beta\ ,\qquad
y(\tau_f)=0\ ,\qquad
u(\tau_f)=0\ ,\qquad
v(\tau_f)=\sqrt2\beta\tanh\psi\ .\hskip15mm\label{jI}
\end{equation}
Since $\tau_f=\tau_e$, the relations (\ref{jIII}) and (\ref{jI}) give the {\it
junction conditions} for geodesics crossing the impulsive wave. It is obvious
that the space coordinates $x$ and $y$ are continuous across the impulsive
hyperboloidal surface $u=0$, whereas the parameter $v$ (along the null rays
generated by the Debever-Penrose vector field ${\bf k}=\partial_v$) suffers
a discontinuity $\Delta v= v(\tau_e)-v(\tau_f)=1/(2\beta^2)$. This behaviour
is in full agreement with a general junction condition for the construction
of non-expanding impulsive waves in Minkowski, de~Sitter and anti--de~Sitter
space-times by the `cut and paste' method \cite{Penrose}, \cite{PG99}.

\section {Conclusions}

We have investigated a class of exact solutions which describe gravitational
and null-matter waves propagating in the anti--de~Sitter universe. By
analyzing geodesic and geodesic deviation, we were able to give
a physical interpretation of these space-times.

We have also demonstrated that these space-times appear non-singular for all geodesic
observers. This is a unique  feature for cosmological waves.
Therefore, the solutions may be considered as an interesting analogue of the
well-known plane gravitational waves in flat Minkowski universe which exhibit
the same property.

Moreover, arbitrary profiles of these waves in the anti--de~Sitter universe can be
prescribed so that sandwich  gravitational plus null-matter waves can easily be obtained.
We have investigated  some of their properties including the geometry of the
wave-surfaces and geodesic motion. This enabled us to construct explicitly impulsive
waves of the Defrise type. The junction conditions across the impulsive hyperboloidal
null surface, which we have derived from the geodesics, are consistent with those
 discussed  in the literature previously.

\section*{Acknowledgments}

I would like to thank Ji\v r\'\i\  Bi\v c\'ak for bringing my attention
to the Siklos class of solutions, and Jerry Griffiths for reading the manuscript.
I also acknowledge the support of grants GACR-202/99/0261 and
GAUK~141/2000 of the Czech Republic and Charles University.

\newpage
{\LARGE Figure Caption}

\bigskip
\noindent
{\bf Figure 1.} The anti--de~Sitter universe, represented as a four-dimensional
hyperboloid in a five-dimensional flat space-time with two time coordinates
$Z_0$ and $Z_4$, is globaly parametrized by the coordinates $T$, $\chi$, $\vartheta$,
$\varphi$. Sandwich waves which propagate in the anti--de~Sitter universe are
bounded by two-dimensional hyperboloidal null surfaces $u=u_1$ and $u=u_2$.
Privileged timelike  geodesics $\psi=const$ in the background are also
indicated.

\bigskip
\noindent
{\bf Figure 2.} The conformal diagram of the anti--de~Sitter universe, with the
global  coordinate chart  $T$, $\chi$, in which $\chi=\pi/2$ represents null
and spacelike infinity. The wave-surfaces $u=const$ of sandwich waves are indicated.
Any timelike observer $\psi=const$ encounters first the sandwich wave propagating
to the left and then the wave propagating in the opposite direction.

\bigskip
\noindent
{\bf Figure 3.} Part of the conformal diagram representing a sandwich wave localized
at $u\in[0, u_2]$. Privileged geodesics $\psi=const$  start at $\tau=0$, $u=-\infty$
at one point in the anti--de~Sitter region~I., given by $u<0$. At $\tau_f$ these
enter the Defrise wave-region~II., and at $\tau_e$ emerge into the anti--de~Sitter
region~III. behind the wave, $u>u_2$.


\begin{thebibliography}{99}
\setlength{\itemsep}{-1mm}


\bibitem{Gowdy}  Gowdy R H 1971
     {\it Phys. Rev. Lett.} {\bf 27} 826

\bibitem{Lukash}  Lukash V N 1975
     {\it Sov. Phys. JETP} {\bf 40} 792

\bibitem{Wain}   Wainwright J 1979
     {\it Phys. Rev.} D {\bf 20} 3031

\bibitem{CCM}    Carmeli M, Charach Ch and Malin S 1981
     {\it Phys. Rep.} {\bf 76} 79

\bibitem{CenMat}   Centrella J and Matzner R A 1982
     {\it Phys. Rev.} D {\bf 25} 930

\bibitem{AHZFLZ} Adams P J, Hellings R W, Zimmerman R L,
    Farhoosh H, Levine D I and Zeldich S
    1982 {\it  Astrophys. J.} {\bf 253} 1

\bibitem{AHZ1}   Adams P J, Hellings R W and Zimmerman R L 1985
     {\it Astrophys. J.} {\bf 288} 14; 1987 {\bf 318} 1

\bibitem{CarVer}   Carr B J and Verdaguer E 1983
     {\it Phys. Rev.} D {\bf 28} 2995

\bibitem{Siklos} Siklos S T C 1985 in
    {\it Galaxies, Axisymmetric Systems and Relativity} ed
    M A H MacCallum (Cambridge: Cambridge University Press)

\bibitem{FeinCha}   Feinstein A and Charach Ch 1986
     {\it Class. Quantum Grav.} {\bf 3} L5

\bibitem{Fein}   Feinstein A  1988
    {\it Gen. Rel. Grav.} {\bf 20} 183

\bibitem{Hogan}  Hogan P A 1988
     {\it Astrophys. J.} {\bf 324} 639

\bibitem{FG}     Feinstein A and Griffiths J B 1994
     {\it Class. Quantum Grav.} {\bf 11} L109

\bibitem{BiGrif}   Bi\v c\' ak J and Griffiths J B 1994
     {\it Phys. Rev.} D {\bf 49} 900

\bibitem{AlGrif1}   Alekseev G A and Griffiths J B 1995
     {\it Phys. Rev.}  D {\bf 52} 4497

\bibitem{AlGrif2}   Alekseev G A and Griffiths J B 1996
     {\it Class. Quantum Grav.} {\bf 13} 2191

\bibitem{BiGrif96}   Bi\v c\' ak J and Griffiths J B 1996
     {\it Ann. Phys. (N.Y.)}  {\bf 252} 180

\bibitem{Berger}   Berger B K, Chru\'sciel P T, Isenberg J and Moncrief V 1997
     {\it Ann. Phys. (N.Y.)}  {\bf 260} 117

\bibitem{Verd}   Verdaguer E 1993
     {\it Phys. Rep.} {\bf 229} 1

\bibitem{BGM}     Bonnor W B, Griffiths J B and MacCallum M A H
    1994  {\it Gen. Rel. Grav.} {\bf 26} 687

\bibitem{JB2000}   Bi\v c\' ak J 2000 in
    {\it Einstein Field Equations and their Physical Implications}
    ed B G Schmidt, Lecture Notes in Physics {\bf 540} (Berlin: Springer Verlag)


\bibitem{Defr}    Defrise L
 1969 {\it Groupes d'isotropie et groupes de stabilit\'e
    conforme dans les escapes lorentziens}
    (Th\'ese: Universit\'e Libre de Bruxelles)

\bibitem{KSMH}     Kramer D, Stephani H, MacCallum M A H and Herlt E
    1980 {\it Exact Solutions of the Einstein's Filed Equations}
    (Cambridge: Cambridge University Press)

\bibitem{Pod98}  Podolsk\'y J 1998
   {\it Class. Quantum Grav.} {\bf 15} 719

\bibitem{BPR}    Bondi H, Pirani F A E and Robinson I 1959
     {\it Proc. Roy. Soc. Lond.}  A {\bf 251} 519

\bibitem{EK}     Ehlers J and Kundt W 1962 in
    {\it Gravitation: an Introduction to Current Research}
    ed L~Witten (New York: Wiley)

\bibitem{BP}    Bondi H and Pirani F A E 1989
     {\it Proc. Roy. Soc. Lond.}  A {\bf 421} 395

\bibitem{BiPo}   Bi\v c\' ak J and Podolsk\'y J 1999
    {\it J. Math. Phys.} {\bf 40} 4506

\bibitem{HE}    Hawking S W and Ellis G F R 1973
    {\it The Large Scale Structure of Space-Time}
    (Cambridge: Cambridge University Press)

\bibitem{ES}     Ellis G F R and Schmidt B G 1977
    {\it Gen. Rel. Grav.} {\bf 8} 915

\bibitem{PG}   Podolsk\'y J and Griffiths J B 1998
     {\it Class. Quantum Grav.} {\bf 15} 453

\bibitem{Pod98a}   Podolsk\'y J 1998
     {\it Class. Quantum Grav.} {\bf 15} 3229

\bibitem{PG97}   Podolsk\'y J and Griffiths J B 1997
     {\it Phys. Rev.} D {\bf 56} 4756

\bibitem{Penrose} Penrose R  1972 in {\it General Relativity}
     ed L~O'Raifeartaigh     (Oxford: Clarendon) p 101

\bibitem{PG99}  Podolsk\'y J and Griffiths J B 1999
     {\it Phys. Lett.} A {\bf 261} 1


\end{thebibliography}
\end{document}